\documentclass[preprint,superscriptaddress,showpacs,nofootinbib,aps]{revtex4}
 \usepackage{graphicx}
 \usepackage{bm}
 \begin{document}
 \title{Study of the ${\eta}^{\prime}$ ${\to}$ $Ve^{+}e^{-}$
        decay with hidden local symmetry model}
 \author{Yueling Yang}
 \email{yangyueling@htu.cn}
 \affiliation{Institute of Particle and Nuclear Physics,
              Henan Normal University, Xinxiang 453007, China}
 \author{Jinshu Huang}
 \email{jshuang@foxmail.com}
 \affiliation{College of Physics and Electronic Engineering,
              Nanyang Normal University, Nanyang 473061, China}
 \author{Gongru Lu}
 \affiliation{Institute of Particle and Nuclear Physics,
              Henan Normal University, Xinxiang 453007, China}
 \begin{abstract}
 Within the hidden local symmetry framework, the Dalitz decay
 ${\eta}^{\prime}$ ${\to}$ $Ve^{+}e^{-}$ is studied with the
 vector meson dominance model. It is found that the partial width
 ${\Gamma}({\eta}^{\prime}{\to}{\omega}e^{+}e^{-})$
 ${\approx}$ $40$ eV and branching ratio
 ${\cal B}({\eta}^{\prime}{\to}{\omega}e^{+}e^{-})$
 ${\approx}$ $2{\times}10^{-4}$, and
 ${\Gamma}({\eta}^{\prime}{\to}{\rho}e^{+}e^{-})$ ${\approx}$
 $10{\Gamma}({\eta}^{\prime}{\to}{\omega}e^{+}e^{-})$
 and ${\cal B}({\eta}^{\prime}{\to}{\rho}e^{+}e^{-})$ ${\approx}$
 $10{\cal B}({\eta}^{\prime}{\to}{\omega}e^{+}e^{-})$.
 The maximum position of the dilepton distribution is
 $m_{e^{+}e^{-}}$ ${\approx}$ $1.33$ MeV.
 These decays are measurable with the advent of high statistics
 ${\eta}^{\prime}$ experiments.
 \end{abstract}
 \maketitle

 \section{Introduction}
 \label{sec01}
 The ${\eta}$ and ${\eta}^{\prime}$ mesons play an important role
 in understanding of the low energy QCD.
 They provide a valuable place for studying three distinct symmetry
 breaking patterns simultaneously (the explicit symmetry breaking
 due to finite quark mass, dynamical spontaneous symmetry breaking,
 and the axial $U(1)$ anomaly) \cite{J.Phys.Conf.Ser.349.012015}.
 In addition, they are the eigenstates of $G$, $C$, $P$, namely,
 $I^{G}J^{PC}$ $=$ $0^{+}0^{-+}$\cite{pdg2012}, so their decays
 are also suitable to test the conservation or breaking of these
 discrete symmetries (such as charge conjugation invariance\footnotemark[1])
 in the strong and electromagnetic interactions
 \cite{Phys.Scr.T99.114,Eur.Phys.J.ST.198.117,J.Phys.Conf.Ser.335.012017}.
 \footnotetext[1]{There are 19 tests of $C$ invariance listed in
 \cite{pdg2012}, including eight of the ${\eta}$ decays
 and six of the ${\eta}^{\prime}$ decays.}

 The ${\eta}^{\prime}$ ${\to}$ $Ve^{+}e^{-}$ decay (where $V$ $=$
 ${\rho}^{0}$, ${\omega}$)\footnotemark[2]
 \footnotetext[2]{The ${\eta}^{\prime}$ ${\to}$ ${\phi}$ transition
 is forbidden by the kinematic constrain.
 The ${\eta}^{\prime}$ ${\to}$ $K^{0{\ast}}e^{+}e^{-}$ decay is a weak
 process, and the permitted weak decays of ${\eta}^{\prime}$ mesons in
 the standard model are expected to occur at the level of $10^{-11}$
 and below \cite{Phys.Scr.T99.104}.}
 is interesting in several respects.
 (1) The ${\eta}^{\prime}$ is the most esoteric meson of the light
 pseudoscalar nonet because it is closely related to the axial $U(1)$
 anomaly of the strong interactions \cite{Int.J.Mod.Phys.E18.1255},
 which is manifested in its heaviest mass and largest width among the
 pseudoscalar nonet \cite{pdg2012}.
 The effect of QCD anomaly should be manifested in the ${\eta}^{\prime}$
 decay modes, besides the ${\eta}$ and ${\pi}^{0}$ decay modes\footnotemark[3].
 \footnotetext[3]{It is well known that all possible strong and electromagnetic
 decays of ${\eta}$ are highly suppressed by various constraints (such
 as $P$, $C$, $G$ parity) \cite{Phys.Scr.T99.114,Eur.Phys.J.ST.198.117}.}
 From the ${\eta}^{\prime}$ ${\to}$ $Ve^{+}e^{-}$ decay, we may
 get some phenomenological implication of the anomaly at low energy.
 (2) Many model-dependent approaches of low energy mesonic interaction,
 e.g. whether the vector meson dominance (VMD) is valid in nature,
 especially, the applicability of the chiral perturbation theory,
 can be tested via ${\eta}^{\prime}$ decays.
 On one hand, the influence of the light vector mesons on the
 ${\eta}^{\prime}$ ${\to}$ $V$ transition form factor and
 branching ratios of ${\eta}^{\prime}$ decay can be investigated;
 On the other hand, the electron-positron invariant mass distribution
 will provide us with some information about the intrinsic structure
 of ${\eta}^{\prime}$ meson and momentum dependence of the transition
 form factor.
 (3) The two-body strong decays of ${\eta}^{\prime}$ ${\to}$
 ${\pi}{\pi}$, ${\pi}{\eta}$ are forbidden by $P$ invariance.
 The electromagnetic decays of ${\eta}^{\prime}$ ${\to}$
 ${\gamma}{\pi}$, ${\gamma}{\eta}$ are forbidden by $C$
 invariance. The main decays of ${\eta}^{\prime}$ meson fall
 into two distinctive classes.
 One class is the hadronic decay into three pseudoscalar mesons
 due to isospin conserving,
 such as ${\eta}^{\prime}$ ${\to}$ ${\eta}{\pi}{\pi}$.
 The other is the radiative decay into particle with quantum
 number $J^{PC}$ $=$ $1^{--}$, such as ${\eta}^{\prime}$
 ${\to}$ ${\rho}{\gamma}$, ${\omega}{\gamma}$, where
 ${\eta}^{\prime}$ ${\to}$ ${\gamma}{\gamma}$ decay is
 the second order electromagnetic transition.
 It is interesting to study the ${\eta}^{\prime}$ ${\to}$
 $Ve^{+}e^{-}$ decay which is related to the two-body
 radiative decay into vector meson and photon, but in this
 case the photon is off-shell.
 To our knowledge, only branching ratio of ${\eta}^{\prime}$ ${\to}$
 ${\omega}e^{+}e^{-}$ decay is estimated to be ${\sim}$ $2{\times}10^{-4}$
 \cite{Phys.Rev.C61.035206} or $(1.69{\pm}0.56){\times}10^{-4}$
 \cite{Eur.Phys.J.A48.190} with the phenomenological VMD model,
 where many resonance parameters are used to fit the data.
 In this paper, the relativistic Breit-Wigner form is
 taken to describe the resonance with the hidden local
 symmetry (HLS) model \cite{hls} which has been tested
 in great detail
 \cite{Eur.Phys.J.C65.211,arxiv:1010.2378,Eur.Phys.J.C55.199,Eur.Phys.J.C73.2453}.
 The measurement about branching ratios of ${\eta}^{\prime}$ ${\to}$
 $Ve^{+}e^{-}$ decay is not given by the Particle Data Group (PDG)
 \cite{pdg2012} because available data on ${\eta}^{\prime}$
 is relatively scarce for the moment.
 However, this situation will be ameliorated with the advent of high
 statistics ${\eta}^{\prime}$ experiments, such as WASA at COSY,
 Crystal Ball at MAMI, BESIII at BEPCII, KLOE-2 at DA${\Phi}$NE,
 and so on \cite{Phys.Rev.D85.014014}.
 There is a necessity to provide a consistent and uniform theoretical
 description for the decay ${\eta}^{\prime}$ ${\to}$ $Ve^{+}e^{-}$.

 \section{theoretical framework and decay amplitudes}
 \label{sec02}
 The HLS model \cite{hls} provides a
 convenient and constraining QCD-inspired framework for studying
 the phenomenology of light mesons in low energy regime of strong
 interactions.
 In this approach, the pseudoscalar mesons are the
 Nambu-Goldstone bosons, and the vector mesons are gauge bosons of
 a spontaneously broken hidden local symmetry that generates their
 Higgs-Kibble masses.
 The anomalous sector (also called WZW \cite{wzw} and FKTUY \cite{fktuy}
 lagrangian) based on HLS allows one to describe the coupling of
 the form $AAP$, $AVP$, $VVP$, $APPP$, and $VPPP$, where $A$, $V$, $P$
 denote the electromagnetic field, vector meson, pseudoscalar,
 respectively.
 The explicit expression of the corresponding lagrangian can be found
 in \cite{Eur.Phys.J.C65.211,arxiv:1010.2378},
 e.g. the triangle anomaly lagrangians can be written as
  \begin{eqnarray}
 {\cal L}_{AAP} &=&
 -\frac{N_{c}e^{2}}{4{\pi}^{2}f_{\pi}}(1-c_{4})
 {\varepsilon}^{{\mu}{\nu}{\alpha}{\beta}}
 {\partial}_{\mu}A_{\nu}{\partial}_{\alpha}A_{\beta}
 {\rm Tr}\big[Q^{2}P\big],
  \label{aap01} \\
 {\cal L}_{AVP} &=&
 -\frac{N_{c}g\,e}{8{\pi}^{2}f_{\pi}}(c_{3}-c_{4})
 {\varepsilon}^{{\mu}{\nu}{\alpha}{\beta}}
 {\partial}_{\mu}A_{\nu}
 {\rm Tr}\big[\{{\partial}_{\alpha}V_{\beta},Q\}P\big],
  \label{avp01} \\
 {\cal L}_{VVP} &=&
 -\frac{N_{c}g^2}{8{\pi}^{2}f_{\pi}}c_{3}
 {\varepsilon}^{{\mu}{\nu}{\alpha}{\beta}}
 {\rm Tr}\big[{\partial}_{\mu}V_{\nu}{\partial}_{\alpha}V_{\beta}P\big],
  \label{vvp01}
  \end{eqnarray}
 where $N_{c}$ $=$ $3$ is the number of color;
 $e^{2}$ $=$ $4{\pi}{\alpha}$;
 $g$ is the universal vector meson coupling constant;
 $f_{\pi}$ ${\approx}$ $92.4$ MeV is the pion decay constant \cite{arxiv:1010.2378};
 $Q$ $=$ ${\rm diag}(2/3,-1/3,-1/3)$ is the quark charge matrix;
 $P$ is the matrix of pseudoscalar meson --- the Goldstone bosons
 associated with the spontaneous breakdown of $G_{\rm global}$ $=$
 $U(3)_{L}{\otimes}U(3)_{R}$;
 and $V$ is the matrix of vector meson --- the gauge bosons of
 the hidden local $U(3)_{V}$ symmetry \cite{Eur.Phys.J.C55.199},
   \begin{equation}
 P=\frac{1}{\sqrt{2}}
   \left( \begin{array}{ccc}
   \displaystyle
   \frac{1}{\sqrt{3}}{\eta}_{0}+\frac{1}{\sqrt{6}}{\eta}_{8}+\frac{1}{\sqrt{2}}{\pi}^{0}
 &{\pi}^{+}
 & K^{+} \\
  {\pi}^{-}
 & \displaystyle
   \frac{1}{\sqrt{3}}{\eta}_{0}+\frac{1}{\sqrt{6}}{\eta}_{8}-\frac{1}{\sqrt{2}}{\pi}^{0}
 & K^{0} \\
   K^{-}
 & \overline{K}^{0}
 & \displaystyle
   \frac{1}{\sqrt{3}}{\eta}_{0}-\frac{2}{\sqrt{6}}{\eta}_{8}
   \end{array} \right),
   \label{p01}
   \end{equation}
   \begin{equation}
 V=\frac{1}{\sqrt{2}}
   \left( \begin{array}{ccc}
   \displaystyle
   \frac{1}{\sqrt{2}}({\omega}+{\rho}^{0})
 &{\rho}^{+}
 & K^{{\ast}+} \\
  {\rho}^{-}
 & \displaystyle
   \frac{1}{\sqrt{2}}({\omega}-{\rho}^{0})
 & K^{{\ast}0} \\
   K^{{\ast}-}
 & \overline{K}^{{\ast}0}
 &{\phi}
   \end{array} \right).
   \label{v01}
   \end{equation}
 The triangle anomaly lagrangians Eqs.(\ref{aap01}---\ref{vvp01})
 depend on two parameters $c_{3}$ and $c_{4}$.
 In our calculation, $c_{3}$ $=$ $c_{4}$ $=$ $1$
 \cite{Eur.Phys.J.C65.211,arxiv:1010.2378}\footnotemark[4],
 \footnotetext[4]{As the statement given in ref.\cite{Eur.Phys.J.C65.211},
 the relation $c_{3}$ $=$ $c_{4}$ $=$ $1$ cannot be considered
 as firmly established without a fully comprehensive fit of all
 relevant measurements. For example, various fits with different
 data sets are presented in ref.\cite{Eur.Phys.J.C65.211}, where one
 global fit with relatively large probability using ND and CMD
 data sample prefers $c_{3}$ $=$ $0.927{\pm}0.010$.}
 so one can obtain the same predictions as the VMD models
 in the triangle anomalous sector,
 i.e. ${\cal L}_{AAP}$ and ${\cal L}_{AVP}$ vanish, and photons
 can only couple to pseudoscalar mesons via the $V$-${\gamma}$
 transitions.

 The physical states ${\eta}$ and ${\eta}^{\prime}$ are mixtures
 of the octet ${\eta}_{8}$ $=$ $\frac{u\bar{u}+d\bar{d}-2s\bar{s}}{\sqrt{6}}$
 and singlet ${\eta}_{0}$ $=$ $\frac{u\bar{u}+d\bar{d}+s\bar{s}}{\sqrt{3}}$
 states.
  \begin{eqnarray}
  \left(\begin{array}{c}
 {\eta}\\ {\eta}^{\prime}
  \end{array}\right) =
  \left(\begin{array}{cc}
 {\cos}{\theta} & -{\sin}{\theta} \\
 {\sin}{\theta} &  {\cos}{\theta}
  \end{array}\right)
  \left(\begin{array}{c}
 {\eta}_{8} \\ {\eta}_{0}
  \end{array}\right),
  \end{eqnarray}
 where ${\theta}$ is the mixing angle,
 with ${\sin}{\theta}$ ${\simeq}$ $-1/3$,
 ${\cos}{\theta}$ ${\simeq}$ $2\sqrt{2}/3$
 \cite{Phys.Scr.T99.55}.

\begin{figure}[ht]
 \includegraphics[width=0.60\textwidth,bb=120 660 490 800]{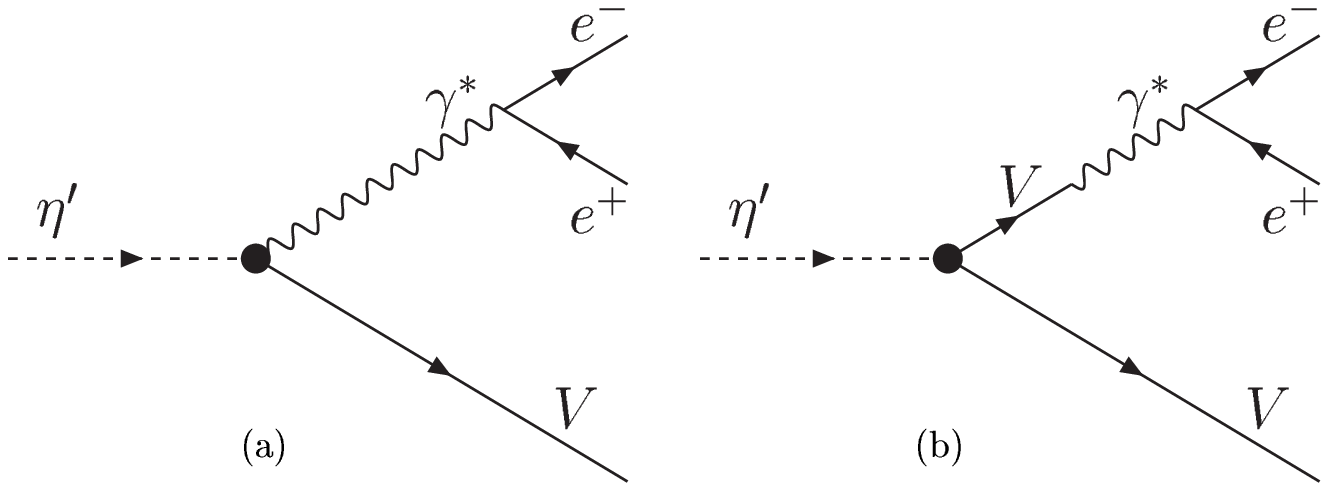}
 \caption{Diagrams contributing to the ${\eta}^{\prime}$ ${\to}$
          $Ve^{+}e^{-}$ decay, where (a) is the direct
          contribution from ${\cal L}_{AVP}$ term, (b) is the VMD
          contribution from ${\cal L}_{VVP}$ term.}
 \label{fig02}
 \end{figure}

 Phenomenologically, the Dalitz decay ${\eta}^{\prime}$ ${\to}$
 $Ve^{+}e^{-}$ is regard as a sequential two-body decays
 chain, i.e. ${\eta}^{\prime}$ ${\to}$ $V{\gamma}^{\ast}$
 ${\to}$ $Ve^{+}e^{-}$.
 In the triangle anomaly lagrangians, terms of ${\cal L}_{AVP}$
 and ${\cal L}_{VVP}$ contribute to decay ${\eta}^{\prime}$
 ${\to}$ $V{\gamma}$.
 The decay amplitude can be written as
   \begin{equation}
  {\cal A}({\eta}^{\prime}{\to}V{\gamma})
  =C_{{\eta}^{\prime}V{\gamma}}
  {\varepsilon}_{{\mu}{\nu}{\alpha}{\beta}}
  p_{\gamma}^{\mu}{\epsilon}_{\gamma}^{\nu}
  p_{V}^{\alpha}{\epsilon}_{V}^{\beta}
  {\times}\big\{ (c_{3}-c_{4})+2c_{3} \big\}
   \label{avp02},
   \end{equation}
 where ${\epsilon}_{\gamma}$ $({\epsilon}_{V})$ and $p_{\gamma}$
 $(p_{V})$ are the polarization vector and four-momentum of
 on-shell photon (vector meson ${\rho}^{0}$ and ${\omega}$),
 respectively;
 the coefficient $C_{{\eta}^{\prime}V{\gamma}}$ contains
 the information of the ${\eta}^{\prime}$ ${\to}$ $V$
 mesonic transition form factor,
   \begin{equation}
  C_{{\eta}^{\prime}{\omega}{\gamma}}
  =\frac{-N_{c}\,g\,e}{48\sqrt{3}{\pi}^{2}f_{\pi}}
   \Big[\frac{f_{\pi}}{f_{8}}{\sin}{\theta}
  +\sqrt{2}\frac{f_{\pi}}{f_{0}}{\cos}{\theta}\Big],
  ~~~~~~
  C_{{\eta}^{\prime}{\rho}{\gamma}}
  =3C_{{\eta}^{\prime}{\omega}{\gamma}}
   \label{avp03},
   \end{equation}
 with the singlet ${\eta}_{0}$ and octet ${\eta}_{8}$ pseudoscalar
 meson decay constant $f_{0}$ ${\simeq}$ $1.04f_{\pi}$ and $f_{8}$
 ${\simeq}$ $1.30f_{\pi}$ \cite{Phys.Scr.T99.55}.
 By the same token, diagrams contributing to the ${\eta}^{\prime}$
 ${\to}$ $V{\gamma}^{\ast}$ ${\to}$ $Ve^{+}e^{-}$
 decay are shown in Fig.\ref{fig02}.
 The corresponding decay amplitude can be written as
   \begin{equation}
  {\cal A}({\eta}^{\prime}{\to}Ve^{+}e^{-})
  =C_{{\eta}^{\prime}V{\gamma}}
  {\varepsilon}_{{\mu}{\nu}{\alpha}{\beta}}
  p_{{\gamma}^{\ast}}^{\mu}
   \frac{\bar{u}_{e^{-}}(-ie{\gamma}^{\nu}){\upsilon}_{e^{+}}}
        {p_{{\gamma}^{\ast}}^{2}+i{\epsilon}}
  p_{V}^{\alpha}{\epsilon}_{V}^{\beta}
  {\times}\bigg\{ (c_{3}-c_{4})+
   \frac{2c_{3}}{1-\frac{p_{{\gamma}^{\ast}}^{2}}{m_{V}^{2}}
            -i\frac{{\Gamma}_{V}}{m_{V}}
           } \bigg\}
   \label{avp04}.
   \end{equation}
 Compared to the two-body decay amplitude Eq.(\ref{avp02}), the
 polarization of the off-shell photon turned into the
 electromagnetic current $j^{\nu}$ $=$
 $\bar{u}_{e^{-}}(-ie{\gamma}^{\nu}){\upsilon}_{e^{+}}$, and the
 VMD factor is dependent on the invariant momentum
 $p_{{\gamma}^{\ast}}^{2}$ $=$ $m_{{\gamma}^{\ast}}^{2}$
 $=$ $(p_{e^{+}}+p_{e^{-}})^{2}$ $=$ $m_{e^{+}e^{-}}^{2}$.

 \section{decay rate and discussion}
 \label{sec03}
 The partial width of the two-body ${\eta}^{\prime}$ ${\to}$
 $V{\gamma}$ decay is
   \begin{equation}
  {\Gamma}({\eta}^{\prime}{\to}V{\gamma})
  =\frac{1}{32{\pi}} \Big(
   \frac{m_{{\eta}^{\prime}}^{2}-m_{V}^{2}}
        {m_{{\eta}^{\prime}}} \Big)^{3}
  {\vert}C_{{\eta}^{\prime}V{\gamma}}{\vert}^{2}
   \label{avp05}.
   \end{equation}
 The differential width of the three-body ${\eta}^{\prime}$ ${\to}$
 $Ve^{+}e^{-}$ decay is
   \begin{equation}
  {\bf d}{\Gamma}({\eta}^{\prime}{\to}Ve^{+}e^{-})
  =\frac{1}{(2{\pi})^{5}}
   \frac{1}{16m_{{\eta}^{\prime}}^{2}}
  {\vert}{\cal A}({\eta}^{\prime}{\to}Ve^{+}e^{-}){\vert}^{2}
  {\vert}\vec{p}^{\ast}_{e^+}{\vert}{\cdot}
  {\vert}\vec{p}_{V}{\vert}
  {\bf d}m_{{\gamma}^{\ast}}
  {\bf d}{\Omega}_{e^{+}}^{\ast}
  {\bf d}{\Omega}_{V}
   \label{avp06},
   \end{equation}
 where (${\vert}\vec{p}^{\ast}_{e^+}{\vert}$, ${\Omega}_{e^{+}}^{\ast}$)
 is the momentum of lepton $e^{+}$ in the rest frame of off-shell
 photon ${\gamma}^{\ast}$;
 and (${\vert}\vec{p}_{V}{\vert}$, ${\Omega}_{V}$) is the momentum of
 vector meson in the rest frame of the decaying ${\eta}^{\prime}$ meson,
   \begin{equation}
  {\vert}\vec{p}^{\ast}_{e^+}{\vert}
  =\frac{{\lambda}^{1/2}(m_{{\gamma}^{\ast}}^{2},m_{e}^{2},m_{e}^{2})}
        {2m_{{\gamma}^{\ast}}}
  =\frac{1}{2}m_{{\gamma}^{\ast}}
   \sqrt{1-\frac{4m_{e}^{2}}{m^{2}_{{\gamma}^{\ast}}}}
  =\frac{1}{2}m_{{\gamma}^{\ast}}{\beta}_{e}
   \label{avp07},
   \end{equation}
   \begin{equation}
  {\vert}\vec{p}_{V}{\vert}
  =\frac{{\lambda}^{1/2}(m_{{\eta}^{\prime}}^{2},m_{{\gamma}^{\ast}}^{2},m_{V}^{2})}
        {2m_{{\eta}^{\prime}}}
   \label{avp08},
   \end{equation}
  where ${\lambda}(a,b,c)$ $=$ $a^{2}+b^{2}+c^{2}-2ab-2bc-2ac$.
  Finally, the differential width in terms of the electron-positron
  invariant mass $m_{e^{+}e^{-}}$ $=$ $m_{{\gamma}^{\ast}}$ can be
  written as
   \begin{equation}
  {\bf d}{\Gamma}({\eta}^{\prime}{\to}Ve^{+}e^{-})
  =\frac{{\alpha}}{96{\pi}^{2}m_{{\gamma}^{\ast}}m_{{\eta}^{\prime}}^{3}}
  {\vert}C_{{\eta}^{\prime}V{\gamma}}{\vert}^{2}
  {\lambda}^{3/2}(m_{{\eta}^{\prime}}^{2},m_{{\gamma}^{\ast}}^{2},m_{V}^{2})
  {\beta}_{e}\big(3-{\beta}_{e}^{2}\big){\bf d}m_{{\gamma}^{\ast}}
   \label{avp09},
   \end{equation}

 With the input parameters collected in TABLE. \ref{tab01}
 (if not specified explicitly, their central values are taken as
 the default input),
 we can get the integrated partial width and the corresponding
 branching ratio of the two-body electromagnetic radiative
 ${\eta}^{\prime}$ decays as follows
   \begin{equation}
  {\Gamma}({\eta}^{\prime}{\to}{\omega}{\gamma})
  =5.426{\pm}0.005_{m_{{\eta}^{\prime}}}
        {\pm}0.010_{m_{\omega}}
        {\pm}0.021_{g}~{\rm keV}
   \label{width01},
   \end{equation}
   \begin{equation}
  {\Gamma}({\eta}^{\prime}{\to}{\rho}{\gamma})
  =54.397{\pm}0.049_{m_{{\eta}^{\prime}}}
         {\pm}0.273_{m_{\rho}}
         {\pm}0.215_{g}~{\rm keV}
   \label{width02},
   \end{equation}
   \begin{equation}
  {\cal B}({\eta}^{\prime}{\to}{\omega}{\gamma})
  =(2.727
  {\pm}0.003_{m_{{\eta}^{\prime}}}
  {\pm}0.005_{m_{\omega}}
  {\pm}0.011_{g}
  {}^{+0.129}_{-0.118}{}_{{\Gamma}_{{\eta}^{\prime}}}
   )\%
   \label{ratio01},
   \end{equation}
   \begin{equation}
  {\cal B}({\eta}^{\prime}{\to}{\rho}{\gamma})
  =(27.335
  {\pm}0.025_{m_{{\eta}^{\prime}}}
  {\pm}0.137_{m_{\rho}}
  {\pm}0.108_{g}
  {}^{+1.295}_{-1.183}{}_{{\Gamma}_{{\eta}^{\prime}}}
   )\%
   \label{ratio02},
   \end{equation}
 where the uncertainties come from $m_{{\eta}^{\prime}}$,
 $m_{V}$, $g$ and ${\Gamma}_{{\eta}^{\prime}}$,
 respectively.
 It is clear that (1) there are two proportions,
 ${\Gamma}({\eta}^{\prime}{\to}{\rho}{\gamma})$ ${\approx}$
 $10{\Gamma}({\eta}^{\prime}{\to}{\omega}{\gamma})$ and
 ${\cal B}({\eta}^{\prime}{\to}{\rho}{\gamma})$ ${\approx}$
 $10{\cal B}({\eta}^{\prime}{\to}{\omega}{\gamma})$,
 due to the Eq.(\ref{avp03}) relationship;
 (2) The largest uncertainty of the predicted branching ratio is
 from the measurement ${\Gamma}_{{\eta}^{\prime}}$;
 (3) These branching ratios are in agreement with the
 measurements ${\cal B}({\eta}^{\prime}{\to}{\omega}{\gamma})$
 $=$ $(2.75{\pm}0.22)\%$ and
 ${\cal B}({\eta}^{\prime}{\to}{\rho}{\gamma})$
 $=$ $(29.3{\pm}0.6)\%$ (including non-resonant
 ${\pi}{\pi}{\gamma}$) \cite{pdg2012}
 within one standard deviation.

 The integrated partial width and the corresponding
 branching ratio of the Dalitz ${\eta}^{\prime}$ ${\to}$
 $Ve^{+}e^{-}$ decays are
   \begin{equation}
  {\Gamma}({\eta}^{\prime}{\to}{\omega}e^{+}e^{-})
  =39.401{\pm}0.040_{m_{{\eta}^{\prime}}}
         {\pm}0.079_{m_{\omega}}
         {\pm}0.156_{g}~{\rm eV}
   \label{width03},
   \end{equation}
   \begin{equation}
  {\Gamma}({\eta}^{\prime}{\to}{\rho}e^{+}e^{-})
  =384.525{\pm}0.377_{m_{{\eta}^{\prime}}}
          {}^{+2.087}_{-2.080}{}_{m_{\rho}}
          {}^{+1.521}_{-1.518}{}_{g}~{\rm eV}
   \label{width04},
   \end{equation}
   \begin{equation}
  {\cal B}({\eta}^{\prime}{\to}{\omega}e^{+}e^{-})
  =(1.980{\pm}0.002_{m_{{\eta}^{\prime}}}
         {\pm}0.004_{m_{\omega}}
         {\pm}0.008_{g}
         {}^{+0.094}_{-0.086}{}_{{\Gamma}_{{\eta}^{\prime}}}){\times}10^{-4}
   \label{ratio03},
   \end{equation}
   \begin{equation}
  {\cal B}({\eta}^{\prime}{\to}{\rho}e^{+}e^{-})
  =( 1.932{\pm}0.002_{m_{{\eta}^{\prime}}}
          {\pm}0.010_{m_{\rho}}
          {\pm}0.008_{g}
          {}^{+0.095}_{-0.087}{}_{{\Gamma}_{{\eta}^{\prime}}}){\times}10^{-3}
   \label{ratio04}.
   \end{equation}
 Similarly, the largest uncertainty of the predicted branching ratio
 comes from the measurement ${\Gamma}_{{\eta}^{\prime}}$.
 Our estimation of the ${\eta}^{\prime}$ ${\to}$ ${\omega}e^{+}e^{-}$
 decay is in good agreement with the previous work
 \cite{Phys.Rev.C61.035206,Eur.Phys.J.A48.190}
 within the uncertainties.
 In ref.\cite{Phys.Rev.C61.035206}, the branching ratio is estimated
 to be about ${\cal B}({\eta}^{\prime}{\to}{\omega}e^{+}e^{-})$
 ${\sim}$ $2{\times}10^{-4}$ with the effective meson theory\footnotemark[5].
 \footnotetext[5]{In ref.\cite{Phys.Rev.C61.035206},
 the multiplicative representation of the transition form factor
 is used to fit the data. That is to say, many vector and/or
 excited vector meson are taken into account. For example, the
 transition form factor $F_{{\omega}{\gamma}{\pi}}$ is written
 as:
 \begin{equation}
 F_{{\omega}{\gamma}{\pi}}(t)
 =\frac{(1+C\,t)m_{\rho}^{2}m_{X}^{2}m_{{\rho}^{{\prime}{\prime}}}^{2}}
       {(m_{\rho}^{2}-t)(m_{X}^{2}-t)(m_{{\rho}^{{\prime}{\prime}}}^{2}-t)}
 \label{formfactor-wrpi},
 \end{equation}
 where the satisfactory fit quality is achieved at the price
 of introducing much more resonance parameters related to the
 corresponding vector mesons.}
 In ref.\cite{Eur.Phys.J.A48.190}, the branching ratio is given
 by ${\cal B}({\eta}^{\prime}{\to}{\omega}e^{+}e^{-})$ $=$
 $(1.69{\pm}0.56){\times}10^{-4}$ with the effective chiral
 Lagrangian\footnotemark[6].
 \footnotetext[6]{In ref.\cite{Eur.Phys.J.A48.190}, the form
 factor of process ${\eta}^{\prime}$ ${\to}$ ${\omega}{\gamma}$
 is written as a function of six parameters (see Eq.(44) and
 Eq.(45) in ref.\cite{Eur.Phys.J.A48.190} for more detail),
 many inputs may cause large uncertainty. \\ \hspace*{3mm}
 In addition, we would like to point out that if the
 parameters $c_{3}$ $\neq$ $1$ in the triangle anomaly lagrangians
 Eq.(\ref{avp01}) and Eq.(\ref{vvp01}), then the partial width
 should be ${\vert}c{\vert}^{2}{\Gamma}({\eta}^{\prime}{\to}V{\gamma})$
 and ${\vert}c{\vert}^{2}{\Gamma}({\eta}^{\prime}{\to}Ve^{+}e^{-})$
 for the ${\eta}^{\prime}$ ${\to}$ $V{\gamma}$ and ${\eta}^{\prime}$
 ${\to}$ $Ve^{+}e^{-}$ decays, respectively.
 For example, if the fitted value of $c_{3}$ $=$
 $0.927{\pm}0.010$ \cite{Eur.Phys.J.C65.211} is used, then the partial
 widthes and branching ratios should be $(85.9{\pm}2.2)\%$ of
 those given in Eq.(\ref{width01})---Eq.(\ref{ratio04}), i.e.,
 the central value of branching ratio of ${\eta}^{\prime}$ ${\to}$
 ${\omega}e^{+}e^{-}$ decay will be $1.70{\times}10^{-4}$, which
 is fairly consistent with that predicted in ref.\cite{Eur.Phys.J.A48.190}.}
 Our results about ${\eta}^{\prime}$ ${\to}$ ${\rho}e^{+}e^{-}$
 decay agree basically with the prediction of
 ${\Gamma}({\eta}^{\prime}{\to}{\pi}^{+}{\pi}^{-}e^{+}e^{-})$
 $=$ $431^{+38}_{-64}$ eV and
 ${\cal B}({\eta}^{\prime}{\to}{\pi}^{+}{\pi}^{-}e^{+}e^{-})$
 $=$ $(2.13^{+0.19}_{-0.32}){\times}10^{-3}$ \cite{Eur.Phys.J.A33.95}
 within the uncertainties, and accord with the recent measurement
 ${\cal B}({\eta}^{\prime}{\to}{\pi}^{+}{\pi}^{-}e^{+}e^{-})$
 $=$ $(2.11{\pm}0.12{\pm}0.14){\times}10^{-3}$ reported by BESIII
 \cite{Phys.Rev.D87.092011} within one standard deviation, where
 almost all of the final states ${\pi}^{+}{\pi}^{-}$ will probably
 come from the resonant ${\rho}^{0}$ meson.
 There are some difficulties for measurement of ${\eta}^{\prime}$
 ${\to}$ $Ve^{+}e^{-}$ experimentally due to the masses of ${\omega}$
 and ${\rho}$ meson are close to each other,
 for ${\eta}^{\prime}$ ${\to}$ ${\omega}e^{+}e^{-}$
 decay, large background might come from ${\eta}^{\prime}$
 ${\to}$ ${\rho}e^{+}e^{-}$, and vice versa.

\begin{figure}[ht]
 \includegraphics[width=0.98\textwidth,bb=50 595 540 680]{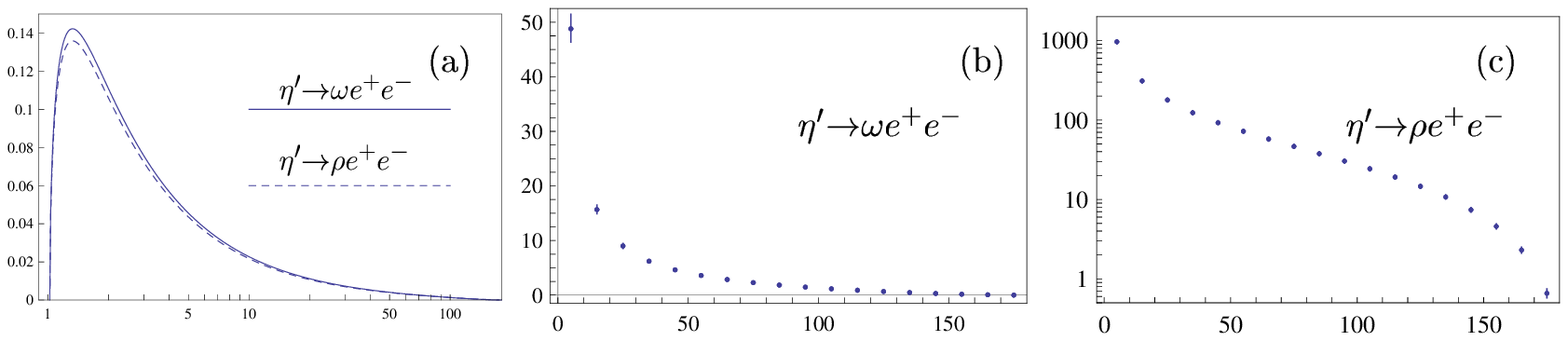}
 \caption{The dilepton spectra of the ${\eta}^{\prime}$ ${\to}$
          $Ve^{+}e^{-}$ decay, where the horizontal axis denotes
          the dilepton invariant mass $m_{e^{+}e^{-}}$ in the unit
          of MeV, and the vertical axis denotes the normalization
          distribution of
          ${\bf d}{\Gamma}({\eta}^{\prime}{\to}Ve^{+}e^{-})/{\bf d}m_{e^{+}e^{-}}$
          in (a) [the area below the line is one],
          some 100 ${\eta}^{\prime}{\to}{\omega}e^{+}e^{-}$
          events in (b) and some 2000 ${\eta}^{\prime}{\to}{\rho}e^{+}e^{-}$
          events in (c).}
 \label{fig04}
 \end{figure}

 Although no available measurement of the ${\eta}^{\prime}$ ${\to}$
 $Ve^{+}e^{-}$ decays is enumerated by PDG so far \cite{pdg2012},
 there is renewed experimental interest in ${\eta}^{\prime}$ decays
 with the advent of high statistics ${\eta}^{\prime}$ experiments.
 For example, some $10^{5}$ fully reconstructed ${\eta}^{\prime}$
 events per day can be reached with WASA at COSY \cite{nucl-ex.0411038};
 Approximately $15{\times}10^{3}$ ${\eta}^{\prime}$ events per hour
 are expected with Crystal Ball at MAMI \cite{Nucl.Phys.PS.B198.174};
 With one year's luminosity at $J/{\psi}$ peak, some 60 million
 ${\eta}^{\prime}$ events could be collected by BESIII at BEPCII
 \cite{J.Phys.G36.085009};
 KLOE-2 at DA${\Phi}$NE experiment expects to increase this sample
 up to about a few $fb^{-1}$ integral luminosity per year within
 the next running \cite{Pizika.B20.221}.
 We take BESIII as an example to estimate the production
 rate of ${\eta}^{\prime}$ ${\to}$ $Ve^{+}e^{-}$ decays.
 It is estimated that there are more than $5{\times}10^{6}$
 ${\eta}^{\prime}$ sample, corresponding to the radiative
 decay $J/{\psi}$ ${\to}$ ${\gamma}{\eta}^{\prime}$ with
 some $10^{9}$ $J/{\psi}$ dataset accumulated at BESIII
 \cite{J.Phys.G36.085009}.
 Given the detection efficiency for ${\eta}^{\prime}$
 ${\to}$ ${\pi}^{+}{\pi}^{-}e^{+}e^{-}$ is about 17\%
 \cite{Phys.Rev.D87.092011}, some 2000
 ${\eta}^{\prime}$ ${\to}$ ${\rho}e^{+}e^{-}$
 and some 100
 ${\eta}^{\prime}$ ${\to}$ ${\omega}e^{+}e^{-}$
 events could be observed at BESIII\footnotemark[7].
 \footnotetext[7]{
  In fact, the ${\omega}$ meson decays mainly into
  $3{\pi}$ and the neutral pion decays mostly into
  two photon. So the detection efficiency will be less
  than 17\% for ${\eta}^{\prime}$ ${\to}$ ${\omega}e^{+}e^{-}$
  decays due to much more final states.
  According to the referee's suggestion, the reconstruction
  efficiency of each photon is ${\sim}$ 80\%, so the
  about 100 events for ${\eta}^{\prime}$ ${\to}$ ${\omega}e^{+}e^{-}$
  could be observed at BESIII.}
 The corresponding distribution of dilepton spectra are
 displayed in Fig.\ref{fig04}.
 Our studies also show that
 (1) the influence of the mass and width of vector mesons on the
 normalization distribution of
 ${\bf d}{\Gamma}({\eta}^{\prime}{\to}Ve^{+}e^{-})/{\bf d}m_{e^{+}e^{-}}$
 is small\footnotemark[8].
 \footnotetext[8]{
  Here, the maximum invariant mass of electron-positron pair is
  $m_{e^{+}e^{-}{\max}}$ $=$ $(m_{{\eta}^{\prime}}-m_{V})$,
  so $p_{{\gamma}^{\ast}{\max}}^{2}/m_{V}^{2}$ $=$
  $m_{{\gamma}^{\ast}{\max}}^{2}/m_{V}^{2}$ $=$
  $m_{e^{+}e^{-}{\max}}^{2}/m_{V}^{2}$ is less than $6\%$.
  In addition, the ratio ${\Gamma}_{V}/m_{V}$ ${\sim}$ 1\% for
  ${\omega}$ meson and ${\sim}$ 20\% for ${\rho}$ meson,
  so with $m_{{\gamma}^{\ast}{\max}}$,
  the factor ${\vert}1/(1-\frac{p_{{\gamma}^{\ast}}^{2}}{m_{V}^{2}}
  -i\frac{{\Gamma}_{V}}{m_{V}}){\vert}^{2}$ ${\approx}$ 1.11
  for ${\omega}$ meson and ${\approx}$ 1.07 for ${\rho}$ meson,
  that is to say, the effects of the mass and width of vector
  meson are about 10\%. At the maximum position in the
  dilepton distribution $m_{e^{+}e^{-}}$ ${\approx}$ $1.33$ MeV,
  the factor ${\vert}1/(1-\frac{p_{{\gamma}^{\ast}}^{2}}{m_{V}^{2}}
  -i\frac{{\Gamma}_{V}}{m_{V}}){\vert}^{2}$ ${\sim}$ 99.99\%
  for ${\omega}$ meson and ${\sim}$ 96.44\% for ${\rho}$ meson,
  that is to say, the effects of the mass and width of vector
  meson are ignorable tiny for ${\omega}$ meson and less than
  4\% for ${\rho}$ meson.}
 (2) The maximum position in the
 distribution is near dilepton threshold\footnotemark[9],
 \footnotetext[9]{
  It is shown from Eq.(\ref{avp09}) that the differential
  width is proportional to $1/m_{{\gamma}^{\ast}}$,
  so the spectra lineshape tends to the maximum with
  $m_{{\gamma}^{\ast}}$ moving to the dilepton threshold.
  In addition, because $m_{{\gamma}^{\ast}}$ is far away
  from the mass of the vector resonance, there is no peak
  near the tail of the dilepton distribution and
  the shapeline is falling down smoothly.}
 i.e. $m_{e^{+}e^{-}}$ ${\approx}$ $1.33$ MeV,
 the corresponding common momentum of vector mesons ${\omega}$
 and ${\rho}$ in the ${\eta}^{\prime}$ rest frame are
 $159.11$ MeV and $164.94$ MeV, respectively.
 This distinctive feature will be helpful to distinguish
 signals from backgrounds.

 \section{summary}
 \label{sec04}
 Based on the triangle anomaly HLS effective lagrangian,
 the interesting ${\eta}^{\prime}$ ${\to}$ $Ve^{+}e^{-}$
 decay is studied with the VDM model.
 Our study show that the partial width
  ${\Gamma}({\eta}^{\prime}{\to}{\omega}e^{+}e^{-})$
  ${\approx}$ $40$ eV and branching ratio
  ${\cal B}({\eta}^{\prime}{\to}{\omega}e^{+}e^{-})$
  ${\approx}$ $2{\times}10^{-4}$, and
  ${\Gamma}({\eta}^{\prime}{\to}{\rho}e^{+}e^{-})$ ${\approx}$
  $10{\Gamma}({\eta}^{\prime}{\to}{\omega}e^{+}e^{-})$
  and ${\cal B}({\eta}^{\prime}{\to}{\rho}e^{+}e^{-})$ ${\approx}$
  $10{\cal B}({\eta}^{\prime}{\to}{\omega}e^{+}e^{-})$,
  which are basically consistent with previous estimation and
  measurement within uncertainties.
  Compared with the radiative decay ${\eta}^{\prime}$ ${\to}$
  $V{\gamma}$, the electron-positron pair, splitting away from
  off-shell photon, its invariant mass is momentum dependent,
  which could provide us with much more information about the
  intrinsic structure of ${\eta}^{\prime}$ meson and form
  factor for electromagnetic transition ${\eta}^{\prime}$
  ${\to}$ $V$.
  It is well known that the charged electron and positron are
  easily identified by the detector saturated with magnetic field.
  In addition, there is distinctive maximum position
  $m_{e^{+}e^{-}}$ ${\approx}$ $1.33$ MeV.
  It can be expected that the era of accurate measurements on
  the ${\eta}^{\prime}$ ${\to}$ $Ve^{+}e^{-}$ decay is coming
  with the advent of high statistics ${\eta}^{\prime}$ experiments.

 \section*{Acknowledgments}
 This work is supported by both {\em National Natural Science Foundation
 of China} under Grant Nos. 11147008, 11275057, U1132103 and U1232101)
 and {\em Program for Science and Technology Innovation Talents in
 Universities of Henan Province} under Grant No. 2012HASTIT030.

  \begin{table}[htb]
  \caption{input parameters for ${\eta}^{\prime}$ ${\to}$ ${\omega}e^{+}e^{-}$ decay}
  \label{tab01}
  \begin{ruledtabular}
  \begin{tabular}{l|cc}
  \multicolumn{1}{c|}{parameter} & value & reference \\ \hline
    mass of ${\eta}^{\prime}$ meson
  & $m_{{\eta}^{\prime}}$ $=$ $957.78{\pm}0.06$ MeV
  & \cite{pdg2012} \\
    mass of ${\omega}$ meson
  & $m_{\omega}$ $=$ $782.65{\pm}0.12$ MeV
  & \cite{pdg2012} \\
    mass of ${\rho}$ meson
  & $m_{\rho}$ $=$ $775.49{\pm}0.34$ MeV
  & \cite{pdg2012} \\
    width of ${\eta}^{\prime}$ meson
  & ${\Gamma}_{{\eta}^{\prime}}$ $=$ $199{\pm}9$ keV
  & \cite{pdg2012} \\
    width of ${\omega}$ meson
  & ${\Gamma}_{\omega}$ $=$ $8.49{\pm}0.08$ MeV
  & \cite{pdg2012} \\
    width of ${\rho}$ meson
  & ${\Gamma}_{\rho}$ $=$ $149.1{\pm}0.8$ MeV
  & \cite{pdg2012} \\
    vector coupling constant
  & $g$ $=$ $5.568{\pm}0.011$
  & \cite{Eur.Phys.J.C65.211}
  \end{tabular}
  \end{ruledtabular}
  \end{table}


 \end{document}